Short Paper

# Filipino Use of Designer and Luxury Perfumes: A Pilot Study of Consumer Behavior

John Paul P. Miranda
Mexico Campus, Don Honorio Ventura State University, Philippines
jppmiranda@dhvsu.edu.ph
(corresponding author)

Maria Anna D. Cruz
Mexico Campus, Don Honorio Ventura State University, Philippines
madcruz@dhvsu.edu.ph

Dina D. Gonzales
Mexico Campus, Don Honorio Ventura State University, Philippines
ddeala@dhvsu.edu.ph

Ma. Rebecca G. Del Rosario
Mexico Campus, Don Honorio Ventura State University, Philippines
mrgdelrosario@dhvsu.edu.ph

Aira May B. Canlas
Mexico Campus, Don Honorio Ventura State University, Philippines
amtbalancio@dhvsu.edu.ph

Joseph Alexander Bansil
Mexico Campus, Don Honorio Ventura State University, Philippines
jabvelayo@dhvsu.edu.ph

**Abstract**

This study investigates the usage patterns and purposes of designer perfumes among Filipino consumers, employing purposive and snowball sampling methods as non-probability sampling techniques. Data was collected using Google Forms, and the majority of respondents purchased full bottles of designer perfumes from retailers, wholesalers, and physical stores, with occasional "blind purchases." Daily usage was common, with respondents applying an average of 5.88 sprays in the morning, favoring fresh scent notes and Eau De Parfum concentration. They tended to alternate perfumes daily, selecting different scent profiles according to the Philippine climate. The study reveals that Filipino respondents primarily use designer perfumes to achieve a pleasant and fresh fragrance. Additionally, these perfumes play a role in boosting self-esteem, elevating mood, and enhancing personal presentation. Some respondents reported fewer common applications, such as using perfume to address insomnia and migraines. Overall, the research highlights the significant role of perfume in the grooming routine of Filipino consumers. This study represents the first attempt to comprehend perfume usage patterns and purposes specifically within the Filipino context. Consequently, its findings are invaluable for manufacturers and marketers targeting the Filipino market, providing insights into consumer preferences and motivations.

## INTRODUCTION

The Philippines is considered one of the fastest-growing luxury markets in Southeast Asia, with the luxury goods sector projected to reach USD 3.3 billion by 2022 (Euromonitor International, 2022; Statista, 2023). Among the various categories of luxury goods, perfumes have emerged as one of the most popular among Filipino consumers (Enriquez, 2017). The increasing interest in luxury and designer perfumes among Filipino consumers has been attributed to rising disposable incomes, changing lifestyles, and a growing desire for self-expression (Dealca et al., 2022). Not to mention that consumers are now, more than ever, exposed to the latest trends and styles not only in fragrances but all luxury items in general through the proliferation of online influencers from various online platforms (Enriquez, 2017). In the case of the Philippine market, Filipinos were considered to have more brand loyalty when compared to their global peers (Lucas, 2013). The country is also considered one of the major consumers of luxury items including perfumes (De Vera, 2021) particularly when they are compared to their Asian counterparts (Estopace, 2010). The willingness of Filipino consumers to spend as well as one of the considered major consumers of luxury items, including perfumes, has been documented in various market research reports (De Vera, 2021; Estopace, 2010).

Therefore, conducting a pilot study on the Filipino usage pattern, and purposes of using designer and luxury perfumes (succeeding referred to as perfumes) is crucial because it can offer insightful information into a field of study that is still largely unexplored, particularly in the context of the Philippines. Understanding Filipino consumers' preferences and usage patterns can help companies create targeted marketing strategies for this lucrative global market. Studying how Filipinos use perfumes can reveal information about their values, beliefs, and sense of self. Perfume use is also frequently linked to cultural and social practices. Last but not least, because perfumes are frequently linked to status and prestige, examining how they are used in the Filipino context can shed light on the socio-economic climate in the country. Thus, a pilot study can fill a gap in the body of knowledge and advance the field's understanding since there is currently no study as far as the literature review conducted by the authors. Therefore, a pilot study on this subject can be a crucial first step in discovering how perfumes are used and perceived in the Filipino market. For these reasons, the study aimed to answer the following questions: 1) What are the respondents' perfume usage patterns, including frequency, time of day, and amount used per spray, as well as their preferences for perfume notes and fragrance concentration? and 2) What are the purposes of using perfumes among the respondents?

## METHODS

This study adopted a pilot descriptive research design, primarily aimed at determining the patterns and reasons for perfume use among adults in the Philippines. The pilot nature of this study was critical for assessing the feasibility of a larger study and identifying potential issues, as highlighted in previous research (Connelly, 2008; Hepburn et al., 2021; Hermans et al., 2022; Hertzog, 2008; Malavenda et al., 2020).

### Sampling Technique

Considering the pilot nature of the study, a smaller sample size was deemed appropriate. The study employed a combination of purposive and snowball sampling methods, both of which fall under non-probability sampling techniques.

### Data Collection

Data were collected using Google Forms. Before distribution, the instrument underwent content validation by experts in the English language. The survey link was disseminated through various social media groups dedicated to perfume enthusiasts, with prior permission obtained from group moderators. The purpose of the survey and the link to the online instrument were posted after receiving consent.

### Survey Administration

The survey was conducted from October to December 2022. It consisted of a 22-item questionnaire divided into three sections: respondent profile, usage patterns, and purchasing purposes. The estimated completion time for the questionnaire ranged from five to fifteen minutes. Due to the use of non-probability sampling methods, the total number of respondents was dependent on the duration of the survey. A total of 88 responses were recorded and used for analysis.

### Data Analysis

Data were analyzed using descriptive and chi-square statistics in the SPSS software. The results of these analyses are presented below.

### Sample Profile

Table 1 shows the profile of the respondents. Statistical analysis showed that most of the respondents, or 47.7%, lived in the National Capital Region (NCR). This observation can be supported by their type of residence, where 92% were residing in an urban area. Higher responses were also recorded from nearby regions (e.g., Central Luzon and CALABARZON). Furthermore, most of the respondents were college graduates (78.4%). This was followed by holders of postgraduate degrees (e.g., Master's or Doctoral degrees) (19.3%) and graduates of either high school or senior high school (2.2%). More than half the respondents were male (56.8%), catholic by religious affiliation (69.3%), and single (78.4%). In addition, around a third of them stated that their family belonged to the middle-income class (₱46,761 - ₱81,832) in the Philippines (Zialcita, 2020). This might be because luxury goods have become more accessible to middle-class consumers (Krishnan et al., 2022; Silva, 2017). These observations were confirmed through one-way $\chi^2$ analysis. It revealed that all were found to have significant differences in their frequency distribution except family income groups (f=26, 29.5%, χ2=18.50, DF=1, p=0.002). This study revealed that majority of them are employed (f=72, 81.8%, $\chi^2$=95.55, DF=2, p<0.001), has college degrees (f=69, 78.4%, $\chi^2$=78.4%, DF=2, p<0.001), dominated by males (f=50, 56.8%, $\chi^2$=27.36, DF=2, p<0.001), catholic (f=61, 69.3%, $\chi^2$=13.14, DF=1, p<0.001), and single (f=69, 78.4%, $\chi^2$=28.41, DF=1, p<0.001).

## Table 1. Respondents Profile

| Variable | *freq* | % | χ² | DF | p-Value |
|---|---|---|---|---|---|
| Age (mean ± SD) | 29.58 ± 7.08 | | - | - | - |
| Residence | | | | | |
| NCR – National Capital Region | 42 | 47.73 | | | |
| Region III – Central Luzon | 15 | 17.05 | | | |
| Region IV-A – CALABARZON | 13 | 14.77 | | | |
| Region VII – Central Visayas | 3 | 3.41 | | | |
| Region X – Northern Mindanao | 3 | 3.41 | | | |
| Region XI – Davao Region | 3 | 3.41 | - | - | - |
| Region II – Cagayan Valley | 2 | 2.27 | | | |
| Region VI – Western Visayas | 2 | 2.27 | | | |
| Region XIII – Caraga | 2 | 2.27 | | | |
| Region I – Ilocos Region | 1 | 1.14 | | | |
| Region V – Bicol Region | 1 | 1.14 | | | |
| Region VIII – Eastern Visayas | 1 | 1.14 | | | |
| Type of residence | | | | | |
| Urban | 81 | 92 | | | |
| Rural | 7 | 8 | - | - | - |
| Employment status | | | | | |
| Employed | 72 | 81.8 | | | |
| Student | 14 | 15.9 | 95.55 | 2 | 0.000 |
| Unemployed | 2 | 2.3 | | | |
| Highest educational attainment | | | | | |
| High School/ Senior High School | 2 | 2.2 | | | |
| College | 69 | 78.4 | 84.30 | 2 | 0.000 |
| Post-graduate | 17 | 19.3 | | | |
| Gender | | | | | |
| Male | 50 | 56.8 | | | |
| Female | 28 | 31.8 | 27.36 | 2 | 0.000 |
| LGBTQIA+ | 10 | 11.4 | | | |
| Religious affiliation | | | | | |
| Catholic | 61 | 69.3 | 13.14 | 1 | 0.000 |
| Non-Catholic | 27 | 30.7 | | | |
| Civil status | | | | | |
| Single | 69 | 78.4 | 28.41 | 1 | 0.000 |
| Married | 19 | 21.6 | | | |
| Family income group* | | | | | |
| Less than ₱11,690 | 0 | 0 | | | |
| ₱11,690 - ₱23,381 | 6 | 6.8 | | | |
| ₱23,381 - ₱46,761 | 19 | 21.6 | | | |
| ₱46,761 - ₱81,832 | 26 | 29.5 | 18.50 | 5 | 0.002 |
| ₱81,832 - ₱140,284 | 17 | 19.3 | | | |
| ₱140,284 - ₱233,806 | 10 | 11.4 | | | |
| At least ₱233,806 | 10 | 11.4 | | | |

## RESULTS

### *Filipino Perfume Usage Patterns and Preferences*

In terms of their first purchase of designer perfume, out of the 88 respondents from the study, 71.6% of them mentioned that their first designer perfume was purchased as a full bottle or brand new in a box (BNIB) (71.6%). While the other 28.4% purchased a designer perfume in the form of a partial bottle or decant, i.e., used perfumes. In addition, the respondents indicated that at one point before the conduct of this study, the maximum number of designer perfumes respondents owned at one point in their life was 300, with an average of 33 perfumes (i.e., full and partial bottles combined). When asked about their method of purchasing designer perfumes, 83% of the respondents often used online resellers from various social media platforms (e.g., FB pages, Instagram accounts, etc.), retailers (72.7%), and physical stores (48.9%) as a way to purchase their designer perfumes. On the other hand, official websites (18.2%) were the least used method. Likewise, regarding the "blind-buying" behavior (i.e., purchasing a perfume without testing it on their skin or knowing how it smells), 35.2% of the respondents stated that they do it sometimes. 28.4% said they often do it, 14.8% either rarely or never, and only 6.8% said they always do it. Furthermore, after seeing a perfume, they tend to wait one month or less before purchasing a BNIB designer perfume (62.2%).

Furthermore, on average, more than a third of the respondents further mentioned that their designer perfumes lasted for over two years (36.4%). This was followed by six months to one year (27.3%) and one to two years (22.7%). Only 13.6% of respondents said their designer perfumes lasted less than six months. Regarding length in using designer perfumes, a third of the respondents had used designer perfumes for more than two and a half years at the time of this study (33%). While only 2.3% of respondents were using them for two years to less than two and half years, 10.2% were using them for one and half years to less than two years, 19.3% were using them for one year to less than one and half years, 20.5% for six months to less than a year, and 14.8% for less than six months. On average, the respondents typically changed or switched to another perfume, mostly daily (47.7%). This was followed by once to thrice a week (34.1%) and three to four times a week (13.6%). In comparison, only 4.5% of respondents mentioned that they changed or switched perfume from five to six times a week.

Table 2 shows the perfume usage of the respondents. The majority of the respondents were using designer perfume daily (f=62, 70.5%) during the start of the day between 5 a.m. to 12 noon (f=66, 75%). This suggests that most of the respondents sprayed perfume in the morning more than any other time of the day. The average sprays by the respondents per usage was 5.88. Further analysis from cross-tabulation suggests that an increase in the usage frequency does not necessarily mean an increase in the average sprays per usage. However, it could be noted that higher average spray usage increased as the frequency of usage also increased for "three to four times a week" and "every day." When the average sprays were multiplied by the frequency of usage, "three to every day" perfume users would spray 5.25 to 44.45 a week.

Table 2. Designer perfume usage

| Perfume usage | *freq* | % | Average sprays per usage | Average sprays per week |
|---|---|---|---|---|
| Usage frequency | | | | |
| Once to twice a week | 7 | 8 | 3.79 | 3.79 − 7.58 |
| Three to four times a week | 16 | 18.2 | 5.25 | 5.25 − 21 |
| Five to six times a week | 3 | 3.4 | 4.17 | 4.17 − 25.02 |
| Everyday | 62 | 70.5 | 6.35 | 6.35 − 44.45 |
| Usage time in a day | | | | |
| Morning | 66 | 75 | 5.94 | - |
| Afternoon | 19 | 21.6 | 5.71 | - |
| Evening | 3 | 3.4 | 5.50 | - |
| TOTAL | 88 | 100 | | |
| Average spray per usage | 5.88 | | | - |

Based on Michael Edwards' fragrance wheel (Donna, 2009), it was revealed from the analysis that most of the respondents ranked perfumes with fresh scent notes as their primary choice of perfume in general (mean = 2.06) and slightly preferred those with floral (mean = 2.06), woody (mean = 2.67), oriental (mean = 2.68). The possible reason for this is that the majority of the respondents from this study were male (Verdugo & Ponce, 2020). Another possible reason for this was due to the season in the Philippines. Further analysis indicates that this can change depending on the climate. For example, when the respondents were asked about their preference according to the climate in the country (i.e., wet and dry), 43.2% of respondents tended to choose perfumes with a woody scent profile during the wet season, followed by oriental (26.1%), floral (19.3%), and fresh (11.4%). More than a third of respondents often choose perfumes with a fresh scent profile (78.4%) during the dry season, followed by floral (14.8%), woody (4.5%), and oriental (2.3%). Moreover, in terms of perfume concentration preference, Table 3 shows that the respondents ranked Eau De Parfum as the top preferred concentration (mean = 2.28). While Eau Fraiche (mean = 4.53) and Aftershave/ mist/ splash were the least preferred by the respondents (mean = 4.55).

Table 3. Preference according to fragrance concentration

| Rank | Concentration | Mean | Verbal Interpretation |
|---|---|---|---|
| 1 | Eau de Parfum | 2.28 | Moderately preferred |
| 2 | Eau de toilette | 2.86 | Slightly preferred |
| 3 | Parfum | 2.99 | Slightly preferred |
| 4 | Eau de cologne | 3.78 | Slightly least preferred |
| 5 | Eau Fraiche | 4.53 | Moderately least preferred |

| 6 | Aftershave/ mist/ splash | 4.55 | Moderately least preferred |

## *Purposes of Using Designer Perfumes*

Table 4 shows the respondents' most commonly cited purposes for using perfumes. The top reason for using designer perfumes for Filipino respondents was "to smell good and fresh," which was given by 88.6% of the respondents. "To gain confidence" (84.1%), "to improve mood" (81.8%), and "to be presentable" (81.8%) are the next three most popular reasons. The respondents also listed the following reasons for using perfumes: "to relax/relieve stress," "to build personal image," "to appear elegant and sophisticated," "to have a signature scent," "to express feelings," and "to get compliments." It also shows that some respondents use designer perfumes for less usual purposes, such as "to treat insomnia" and "to cure headaches."

Table 4. Purposes of using designer perfumes

| Rank | Purposes | freq | % |
|---|---|---|---|
| 1 | To smell fresh and good | 78 | 88.6% |
| 2 | To gain confidence | 74 | 84.1% |
| 3 | To improve my mood | 72 | 81.8% |
| 4 | To be presentable | 72 | 81.8% |
| 5 | To relax/ relieve stress | 61 | 69.3% |
| 6 | To build a personal image | 59 | 67% |
| 7 | To appear elegant and sophisticated | 50 | 56.8% |
| 8 | To have a signature scent | 49 | 55.7% |
| 9 | To express my feelings | 45 | 51.1% |
| 10 | To get compliments | 43 | 48.9% |
| 11 | To supercharge positivity | 39 | 44.3% |
| 12 | To stand out from the crowd | 38 | 43.2% |
| 13 | To trigger memories | 37 | 42% |
| 14 | To allure attractiveness | 32 | 36.4% |
| 15 | To be more daring | 17 | 19.3% |
| 16 | To be popular | 6 | 6.68% |
| 17 | To treat insomnia | 3 | 3.4% |
| 18 | To cure headache | 2 | 2.3% |

## DISCUSSION

The high concentration of respondents in urban areas, particularly in the National Capital Region, is in line with worldwide trends showing higher luxury product consumption in metropolitan areas (Dobbs et al., 2012; Cham et al., 2017; Islam & Singh, 2020; Mahalder & Rahman, 2020). This suggests that urban residents have better access to and awareness of luxury items, including perfumes. The substantial proportion of college-educated respondents points to a potential link between higher education and luxury item interest, likely due to enhanced consumer knowledge and financial capability linked to higher educational levels. Additionally, the gender distribution among respondents, primarily male, contradicts the traditional view of perfume as a predominantly female product. This indicates shifting societal norms and growing acceptance of fragrance use among men. The demographic lean towards single and Catholic respondents may suggest specific cultural or social influences on perfume purchasing patterns in the Philippines.

Another key aspect of the study is the uniformity of perfume usage across different family income groups, challenging the notion that luxury goods are exclusive to the wealthier classes. This potentially indicates a democratization of luxury goods consumption in the Philippines. In terms of buying behavior, a significant majority of respondents prefer purchasing full, brand-new bottles of designer perfumes. This trend may suggest a consumer desire for authenticity and a complete buying experience. The prominent use of online resellers and social media platforms for perfume purchases underlines the increasing significance of digital marketplaces in the luxury goods sector. At the same time, this study provided some insights into the scent preferences of Filipino consumers, with a marked preference for fresh scent notes. This inclination could be influenced by gender differences and the climate of the country, as variations in scent preferences were observed between the wet and dry seasons. Moreover, the preference for Eau De Parfum over other perfume concentrations suggests a consumer desire for a balance between scent intensity and longevity.

Respondents indicated various reasons for using perfumes, such as wanting to smell good and fresh, gaining confidence, improving mood, and presenting themselves well. These reasons highlight the role of fragrances in enhancing personal image and emotional well-being. The use of perfumes for less conventional purposes, like treating insomnia and curing headaches, indicates an overlap between cosmetic and medicinal uses of perfumes, pointing to an evolving relationship between consumers and fragrances. This result is partially supported by earlier studies that suggest aromatherapies are sometimes used to relieve several ailments (Ali et al., 2015; Mandal et al., 2021; White & Downers, 2011). Additionally, the results from this study also imply that the majority of respondents view designer perfumes as essential for maintaining personal hygiene, with the need to smell pleasant and fresh ranking highest. The results also show that respondents use perfumes for a variety of other purposes, including self-confidence, mood enhancement, and appearance (Ribeiro, 2017; Zhang et al., 2019). This shows that many Filipino consumers might use designer perfumes to boost their self-esteem and general well-being in addition to being a cosmetic product (Ribeiro, 2017). It is interesting to note that several respondents stated less typical perfume uses, such as relieving migraines and treating insomnia, suggesting that Filipino customers may consider designer perfumes to have medicinal advantages. Additionally, these results offer insightful information into the purposes behind Filipino consumers of perfumes, which might be used to design marketing plans and merchandise specifically tailored to their wants and needs.

## CONCLUSION, RECOMMENDATIONS, AND LIMITATIONS

The study provides insights into the perfume usage patterns of the respondents. The majority of respondents bought their first bottle of designer fragrance in full, and they frequently bought perfumes from physical stores, online retailers, and resellers. The respondents had a high prevalence of blind-buying behavior, and they typically waited one month or less before buying a BNIB designer perfume. The respondents changed or switched to a different designer fragrance daily, and they had been using them for an average of more than two and a half years. The top preferred concentration of Eau De Parfum was chosen by the respondents, who, in general, preferred perfumes with fresh scent notes. This study also shows that the respondents sprayed perfume more frequently in the morning than at any other time of the day. It also shows that an increase in usage frequency does not always translate into an increase in the average number of sprays per usage.

Furthermore, according to respondents, the most frequently cited purpose for using perfumes is to smell pleasant and fresh. Designer scents are viewed as a way for Filipino customers to maintain good personal hygiene and a pleasing aroma. It can also be concluded from the results that there may be additional reasons for using scents, such as improving confidence, mood, and looks. It is amusing to observe that some individuals use perfumes for uncommon purposes like managing sleeplessness and headaches, which raises the possibility that using designer perfumes may have therapeutic benefits. Likewise, this study also offers insightful information about the preferences and requirements of Filipino customers for perfumes, which can help businesses create their marketing plans and product lines to satisfy these needs.

As for the recommendations, perfume companies should concentrate their marketing efforts on these scents in light of the study's findings. Since most respondents spray perfume in the morning, businesses might want to emphasize in their marketing campaigns the advantages of doing so to penetrate the Filipino market. When designing their products, businesses should consider the number of sprays used per application and the frequency of application. These suggestions can help designer perfume companies enhance their marketing plans and product development to better cater to the wants and preferences of the Filipino market.

Future research should look into the limitations of this study. For instance, there are two specific limitations identified. First, the data collection methodology used in this study. The use of online surveys through social media following the combined implementation of purposive and sampling techniques. The use of this kind of data collection may have restricted the ability of this study to investigate the specific aspects of the phenomenon in this study. Second, the small number of respondents may affect the results and insights provided by this study. For this reason, it is recommended that other data collection methods and increasing the number of respondents should be considered in future research undertakings to understand better and provide more thorough insights about the topic. In addition to these limitations is its small sample size of 88 respondents, which is normal for a pilot study. However, there are some benefits to this as well, including the capacity to collect data more thoroughly and execute the study swiftly and affordably. Although the preliminary findings of this pilot study should be viewed cautiously, they will be used to enhance the larger study's design. A bigger sample size and more generalizable results are possible with the larger study. Although a drawback, the small sample size is a must in the research process to guarantee the viability and caliber of the larger study.

## PRACTICAL IMPLICATIONS

This study offers fresh insights for Asian businesses, particularly in the Philippine perfume industry which can guide manufacturers, sellers, and marketers in strategic decision-making. Manufacturers are advised to focus on developing perfumes with fresh scent notes and Eau de Darfum concentrations if they want to specifically to Filipino consumer preferences. Additionally, the creation of climate-adapted perfumes and culturally resonant packaging can significantly enhance product appeal. For marketers, emphasizing the mood-enhancing and self-esteem-boosting benefits of designer perfumes in advertising campaigns is recommended, aligning with the identified motivations of Filipino consumers. Furthermore, maintaining a robust presence in physical retail spaces is vital due to their significant influence on Filipino purchasing behaviors. However, exploring online channels for potential "blind purchases" is also essential, considering the growing digital trend. Sellers can enhance customer engagement through educational initiatives about varied perfume uses and interactive events in physical stores, alongside leveraging social media for community building. Overall, the insights from this study can be valuable in shaping product development, marketing strategies, and customer relations which in turn can help businesses to navigate and successfully cater to the diverse preferences of the Filipino perfume market.

## ACKNOWLEDGEMENT

The authors of this study extend their sincere gratitude to all the respondents who contributed their valuable time to this research.

## FUNDING

The study did not receive funding from any institution.

## DECLARATIONS
### Conflict of Interest

The authors hereby declare that there is no conflict of interest associated with this study.

## Informed Consent

Informed consent was obtained from all respondents involved in this study. Consent declaration was integrated into the Google Forms used for data collection, ensuring that respondents were fully informed about the nature of the study and their involvement in it.

## Ethics Approval

Ethics approval was not obtained for this study as it was determined not to be required. However, the study strictly adhered to the guidelines set by the Philippine Health Research Ethics Board (PHREB) for conducting research. Additionally, all procedures complied with the data privacy laws of the Philippines, ensuring the protection of the respondent's personal information.

## Author's Biography

John Paul P. Miranda is an associate professor at Don Honorio Ventura State University – Mexico Campus and concurrently serves as the project head for international linkages and partnerships at the Office for International Partnerships and Programs at the same institution.

Maria Anna D. Cruz is the current Area Chairperson for the College of Industrial Technology and Hospitality Management at Don Honorio Ventura State University – Mexico Campus.

Dina D. Gonzales is a faculty-researcher at Don Honorio Ventura State University – Mexico Campus.

Ma. Rebecca G. Del Rosario is a faculty-researcher at Don Honorio Ventura State University – Mexico Campus.

Aira May B. Canlas is the current Area Chairperson for the College of Business Studies and Accountancy at Don Honorio Ventura State University – Mexico Campus.

Joseph Alexander Bansil is a faculty-researcher at Don Honorio Ventura State University – Mexico Campus.